\documentclass[conference]{IEEEtran}

\usepackage{url}

\usepackage{breakurl}

\usepackage{fancyhdr}
\usepackage{graphicx}
\usepackage{authblk}
\usepackage{xcolor}
\usepackage{verbatim}
\usepackage{subcaption}

\def\BibTeX{{\rm B\kern-.05em{\sc i\kern-.025em b}\kern-.08em
    T\kern-.1667em\lower.7ex\hbox{E}\kern-.125emX}}

\usepackage{amsmath}
\usepackage{algorithm}
\usepackage[noend]{algorithmic}
\algsetup{linenosize=\tiny}
\newcommand{\llIf}[2]{{\let\par\relax\lIf{#1}{#2}}}
\newcommand{\llElse}[1]{{\let\par\relax\lElse{#1}}}

\usepackage{makecell}
\usepackage{multirow}

\newcommand{\svac}[1]{}
\newcommand{\svab}[1]{}

\def\sva#1{}
\def\john#1{}

\def\hl#1{#1}
\def\johnhlb#1{{#1}}
\def\johnhl#1{{#1}}
\def\asplos21hl#1{#1}
\def\iscahl#1{#1}

\def\microhl#1{{#1}}
\def\iscarhl#1{{#1}}
\def\iscarm#1{}

\begin{document}
%

\title{A Case for Fine-grain Coherence Specialization in Heterogeneous Systems
} 


\date{}
\author[1]{Johnathan Alsop}
\author[2]{Weon Taek Na}
\author[3]{Matthew D. Sinclair}
\author[2]{Samuel Grayson}
\author[2]{Sarita V. Adve} 

\affil[1]{AMD Research}
\affil[2]{University of Illinois at Urbana-Champaign} 
\affil[3]{University of Wisconsin - Madison}

%
%
%

%
%


\maketitle





%

\begin{abstract}

Hardware specialization is becoming a key enabler of energy-efficient performance.
Future systems will be increasingly heterogeneous, integrating multiple specialized and programmable accelerators, each with different memory demands.
Traditionally, communication between accelerators has been inefficient, typically orchestrated through explicit DMA transfers between different address spaces.
More recently, industry has proposed unified coherent memory which enables implicit data movement and more data reuse, 
but often these interfaces limit the coherence flexibility available to heterogeneous systems.

This paper demonstrates the benefits of fine-grained coherence specialization for heterogeneous systems.
We propose an architecture that enables low-complexity independent specialization of each individual coherence request in heterogeneous workloads by building upon a simple and flexible baseline coherence interface, Spandex.
We then describe how to optimize individual memory requests to improve cache reuse and performance-critical memory latency in emerging heterogeneous workloads. 
Collectively, our techniques enable significant gains, reducing execution time by up to 61\% or network traffic by up to 99\% while adding minimal complexity to the Spandex protocol.

\end{abstract}

\vspace{-1ex}
\section{Introduction}
\label{sec:intro}
\vspace{-1ex}

In nearly all compute domains, hardware specialization and parallelism have become important drivers of compute performance.
GPUs, FPGAs, and other specialized devices are being incorporated into systems ranging from mobile and IoT devices to supercomputers and datacenters.
However, the conventional SoC accelerator integration process is inefficient and complex.
Communication with a new accelerator often uses explicit copies to and from DRAM.
This enables simple integration and the use of specialized memories for each accelerator, but it prevents implicit reuse, limits programmability, and makes inter-device communication inefficient.
As accelerators become a more important part of emerging applications and the memory wall increases the importance of efficient cache reuse,
coherence specialization is becoming just as important to performance as compute specialization. 


\iscahl{
In order to efficiently adapt to emerging workloads and systems, mechanisms for coherence optimization must be both flexible and simple.
There have been many coherence extensions proposed over the years (discussed further in Section~\ref{sec:related}), but these generally build upon conventional hardware protocols originally designed for CPUs such as MESI.
Such protocols are effective for a wide range of CPU workloads, but these complex coherence strategies often incur unacceptable overheads for accelerators such as GPUs~\cite{HechtmanChe2014, HechtmanSorin2013, SinghShriraman2013}.
In addition, the complexity of MESI-based protocols makes validating protocol changes expensive, requiring that the cost of any coherence extension be amortized over a broad range of general-purpose applications.
A simpler baseline protocol would make it practical to validate coherence extensions for an individual application domain or class of accelerators.
}
Recently, the Spandex protocol was proposed as a simple and flexible alternative to MESI-based coherence for heterogeneous systems~\cite{AlsopSinclair2018}.
Spandex is based on the 
DeNovo protocol~\cite{ChoiKomuravelli2011}, and is designed to simply and flexibly interface a diverse set of devices (CPU cores, GPUs, and accelerators), each with different static coherence strategies (e.g., MESI, GPU, or DeNovo coherence protocols). Spandex achieves this by identifying an interface consisting of a set of coherence request types that represent the diversity of the above protocols.
Spandex avoids many of the pitfalls of a complex MESI-based approach, but it misses out on an entire dimension of coherence flexibility: the ability to mix and match specialized coherence request types from a device based on the properties of each individual access of that device. 

In this work we advocate for fine-grain coherence flexibility in heterogeneous memory systems.
By specializing the request type used for every access individually rather than at device granularity, we can motivate new coherence specializations that increase cache reuse, reduce data movement, and favor critical accesses over non-critical accesses.
For example, forwarding updates directly from a producer device (writer) to a consumer device (reader) can improve performance for some workloads.
However, this kind of data forwarding is not applicable for all writes. 
We therefore propose a trace-based technique for automatically determining the best specialized request type to be used for each memory access, and we show that this can improve cache efficiency relative to device granularity coherence specialization for common access patterns.
In addition, we show that adding such specializations to a coherence protocol is not always a complex and costly process; protocol complexity can be controlled as long as a simple and flexible baseline protocol like Spandex is used.
Overall, we 
make the following contributions:

\begin{itemize}

  
   \item We propose specializing coherence request types at the granularity of an individual access in a heterogeneous application, rather than at the granularity of a device (such as CPU, GPU, and other accelerators). As our initial fine-grain coherence specialized system, we enable, within a device, all request types supported by Spandex.

  \item 
  We describe two opportunities for further coherence specialization that can be used to improve cache performance for certain access patterns when fine-grained coherence specialization is enabled. 
  
  \item We quantitatively show that adding the new specializations on top of Spandex is 2$\times$ less complex than adding them to a more conventional MESI-based protocol.

  \item To efficiently leverage our fine-grained coherence specializations, we present an algorithm that selects request types for individual accesses in a heterogeneous application to maximize cache cache efficiency and performance.


  
  \item 
  We evaluate fine-grain specialized coherence on a range of heterogeneous workloads with diverse access patterns and show that it can reduce execution time by up to 61\% and network traffic by up to 99\%, while also motivating programming patterns for heterogeneous systems that would otherwise be inefficient. 
\end{itemize}
By combining fine-grain coherence specialization with software that has been co-designed to exploit these new opportunities, we show that it is possible to significantly increase cache reuse, reduce data movement, and improve performance in heterogeneous applications, while keeping design complexity under control.

\vspace{-1ex}
\section{Related Work}
\label{sec:related}
\vspace{-1ex}


There are multiple active efforts to design coherence for heterogeneous systems~\cite{OpenCAPISpec,GenZSpec,CCIXSpec,CHISpec,CXLSpec2019,MungerAkeson2016}.
Many of these offer a high degree of flexibility, enabling ``non-snoopable" regions and request types (similar to self-invalidated loads and write-through stores).
However, in many cases these properties are fixed for a given address region, preventing the use of different request types for a given address at different times or from different devices.
Some interfaces also offer mechanisms for pushing writes directly to a designated node  (e.g., cache stashing in ARM CHI), similar to write-through forwarding optimizations described in Section~\ref{sec:design}.
However, since these mechanisms may update coherence state for the target data, they incur added complexity in the form of transient states, and they require indirection through the LLC (direct communication is not possible).
In general, these emerging coherence interfaces are not focused on request type optimization at the level of individual accesses, so they miss opportunities for coherence flexibility.
Past work has studied the benefits of adaptive
cache policies~\cite{LaiFalsafi2000,LebeckWood1995,NilssonStenstrom1994}
and of flexible cache state or coherence granularities~\cite{CantinSmith2006,ChoiKomuravelli2011,DavariRos2015,KumarZhao2012,RothmanSmith2000,TorrellasLam1994,TrancosoTorrellas1996,ZhaoShriraman2013}. 
Protocol extensions have also been proposed to detect and predict specific sharing patterns, optimizing requests to reduce wasteful traffic and latency for those patterns~\cite{AbdelShafiHall1997,AcacioGonzalez2002,AndersonKarlin1996,BilirDickson1999,ChengCarter2007,CoxFowler1993,HossainHemayet2008,KaxirasGoodman1999,KaxirasKeramidas2010,KoufatyChen1996,KurianKhan2013,MartinHarper2003,MartinHill2003,MukherjeeHill1998,PoulsenYew1994,RamachandranShah1995,RosAcacio2008,SomogyiWenisch2009,StenstromBrorsson1993}.
All of the above innovations help to motivate and demonstrate the benefits of dynamic adaptability, 
and the methods they propose for detecting and predicting sharing patterns will be increasingly relevant to specialized memory systems. 
However, these innovations focus on optimizing communication patterns in the context of multicore CPU workloads rather than heterogeneous workloads, 
and the benefits they provide often come at the cost of additional overhead and complexity on top of an already complex MESI-based protocol.

\iscahl{
A wide range of past work explores 
allowing software to more explicitly control data movement and data locality 
(\cite{BeckmannSanchez2013,Gelernter1989,HardavellasFerdman2009,HoffmannWentzlaff2010,HossainDwarkadas2011,LeeKim2011,YelickBonachea2007}). These data movement optimizations tend to involve simple hardware changes which minimally affect the coherence protocol, but they do not provide the dynamic flexibility enabled by coherence specialization at the granularity of individual accesses. 
}


Many studies also have proposed ways to improve memory access efficiency through adaptive cache management (e.g., replacement, prefetching, and bypassing) in both CPUs and GPUs based on hints from the software, compiler, or runtime profilers~\cite{AyersLitz2020,BeylsDhollander2005,ChenChang2014,KooOh2017,LiangXie2015,PapaefstathiouKatevenis2013,SethiaDasika2013,TianPuthoor2015,WangMcKinley2002,YangGovindarajan2005}.
These techniques allow caches to adapt to different access patterns and improve cache efficiency, and many of the insights regarding cache policy prediction can be leveraged to help select request types in a system with fine-grain coherence specialization.
However, their flexibility is constrained to cache management policies rather than coherence protocol flexibility. 

Recent work proposes communicating high level properties of a program's memory access pattern to the underlying hardware to guide architectural optimizations such as cache policy specialization, intelligent data mapping in DRAM and intelligent GPU work distribution~\cite{VijaykumarEbrahimi2018, VijaykumarJain2018}. 
These techniques are compatible with a Spandex system, and could take advantage of fine-grain coherence specialization by guiding request type selection.

\johnhl{Recent work has improved utilization, load balance, and data reuse by optimizing the scheduling and mapping computation tasks to compute devices in dataflow workloads.
These efforts show significant gains for a range of CPU~\cite{ZhangRajbhandar2018} and GPU/accelerator~\cite{AilaLaine2009, LuYan2017, SteinbergerKenzel2014, ZhengOh2017} applications.}
These software frameworks rely on an awareness of reuse potential, parallelism, data dependencies, and producer-consumer relationships to enable more data-efficient compute.
This awareness makes them well suited to exploit fine-grain coherence flexibility to further improve reuse or reduce wasteful communication overheads in emerging workloads.

\johnhlb{Some recent work identifies the optimal cache coherence policy for different combinations of composable accelerators, easing the process of heterogeneous system design~\cite{BhardwajHavasi2020,MantovaniGiri2020}.
However, flexibility is limited to a single coherence decision for each kernel, limiting the types of possible optimizations.}


Overall, the adaptivity innovations of past work complement the insights of this work, and can be used to guide request type decisions and coherence extensions 
for future heterogeneous systems and workloads.

\vspace{-1ex}
\section{Background: Heterogeneous Coherence}
\label{sec:back}
\vspace{-1ex}

Devices in heterogeneous systems can have very different memory demands which motivate very different coherence strategies.
Table~\ref{table:spandex_reqs} classifies coherence request types used by MESI (used for latency-sensitive CPUs), GPU coherence (used for throughput-sensitive GPUs), and DeNovo (a hybrid protocol designed for CPUs and GPUs~\cite{ChoiKomuravelli2011,SinclairAlsop2015}). 
It also includes additional protocol flexibility proposed by this work in bold (these new request types are described in Section~\ref{sec:design-optimizations}).
The Spandex coherence interface was proposed as a way to integrate all three of the above coherence strategies in a heterogeneous system, and therefore supports all (non-bolded) request types.
\johnhlb{Although the following discussion considers a system with a shared last level cache (LLC), the same tradeoffs apply to a stateless LLC.}


\begin{table}
\centering
{\footnotesize
\begin{tabular}{| c | c | c | c |} 

\hline
\textbf{\makecell{Coherence \\ Dimension}} & \textbf{\makecell{Options}} & \textbf{\makecell{Request \\ Type}}  & \textbf{\makecell{Used For}} \\ \hline

\multirow{3}{*}{\makecell{Stale Data \\ Invalidation}} & \multirow{2}{*}{\makecell{Self-inval.}} & ReqV & \makecell{DeNovo (LD) \\ GPUc (LD) } \\ \cline{3-4}

 & & \textbf{ReqVo} & \textbf{FCS(LD)} \\ \cline{2-4}

 & \makecell{Writer-inval.} & ReqS & \makecell{MESI (LD)} \\ 
 \hline

 \multirow{5}{*}{\makecell{Update \\ Propagation}} & \multirow{2}{*}{Ownership} & ReqO  & DeNovo (ST)\\ \cline{3-4}

 & & ReqO+data & \makecell{MESI (ST,RMW) \\ DeNovo (RMW) \\ \textbf{FCS(LD)}} \\ \cline{2-4}

 & \multirow{3}{*}{\makecell{Write- \\ through}} & ReqWT\textbf{[fwd]} & \makecell{GPUc (ST) \\ \textbf{FCS(ST)}} \\ \cline{3-4}

 & & \textbf{ReqWTo} & \makecell{\textbf{FCS(ST)}} \\ \cline{3-4}

 & & ReqWT\textbf{[fwd]}+data & \makecell{GPUc (RMW) \\ \textbf{FCS(RMW)}} \\ \cline{3-4}

 & & \textbf{ReqWTo+data} & \textbf{\makecell{FCS (RMW)}} \\ \hline

\multirow{2}{*}{\makecell{Request \\ Granularity}} & Word & all & \makecell{DeNovo (all) \\ GPUc (ST,RMW)} \\ \cline{2-4}

 & Line & all & \makecell{MESI (all) \\ GPUc (LD)} \\ \hline

\end{tabular}
\caption{Coherence request type classification and what they are used for in a static DeNovo, GPU coherence (GPUc), or MESI protocol. Bolded text represents coherence flexibility added by fine-grain coherence specialization (FCS).} 

\label{table:spandex_reqs}
\vspace{-2ex}
}
\end{table}

\noindent
\textbf{Stale Data Invalidation}:
Spandex loads can generate self-invalidated reads (ReqV) or writer-invalidated reads (ReqS).
Self-invalidated reads do not obtain any permissions 
and do not change coherence state at the LLC or other caches, but they require the requesting device to self-invalidate the data's local private copy before it can be read at a later time in a stale state (in a data race-free, or DRF, system this can be achieved by self-invalidating all valid data in the private cache at synchronization points).
This simple invalidation mechanism can enable higher throughput scalability at low energy and latency costs, and thus it is used for loads in both DeNovo and GPU coherence. 

In contrast, CPU coherence protocols like MESI use writer-invalidated reads (ReqS). 
This means that reads obtain sharer permissions, and writes must revoke these permissions from any sharer caches.
Therefore MESI-style protocols allow readers to cache data until it is known to be stale (no self-invalidation needed).



\noindent
\textbf{Update Propagation}:
Stores and atomic operations in a Spandex system can use ownership-based updates (ReqO and ReqO+data) or write-through updates (ReqWT and ReqWT+data).
Obtaining ownership for an update enables local reuse if locality exists for that address, but it also adds latency for subsequent requests from remote devices because the ID of the owning device must be queried at the LLC and the request must then be forwarded to the owner before it can be handled.
Both MESI and DeNovo obtain ownership for stores and atomics, enabling high cache reuse for modified data.

Write-through requests flush updates to the LLC without obtaining exclusive permission.
This ensures that the up-to-date data will be present at the LLC for future remote accesses, but it limits the reuse potential at the updating device.
GPU coherence uses write-through requests for updates, avoiding the overheads of ownership tracking but trading off reuse opportunity at the updating cache.

\noindent
\textbf{Request Granularity}:
Each of the above request types may use word or line granularity.
MESI typically uses line granularity for all access types.
This is effective at exploiting spatial locality and mitigates the high overheads of MESI state tracking, but it can incur false sharing, and it may require requesting both ownership and up-to-date data (ReqO+data) on a store miss (if the unwritten words in the line are not valid).
GPU coherence use line-granularity requests for loads and word granularity requests for stores, while
DeNovo uses word-granularity for both load and store requests (although load responses will be at line granularity if the data is available at the responder).
As a result, both DeNovo and GPU can exploit spatial locality for loads, but avoid false sharing overheads and do not need to request the up-to-date data on a store miss.

\vspace{-1ex}
\section{Fine-grain Coherence Specialization}
\label{sec:design}

\subsection{Mixing and Matching Coherence Request Types}

While Spandex is an efficient interface for enabling coherence specialization at device granularity,
in order to further optimize data movement in emerging systems we propose specializing coherence at the granularity of an individual request.
By choosing a request type for each access individually, software can incur the overheads of owner permissions, sharer permissions, and coarse-grain data movement only when it makes sense for the workload.
In addition, this flexibility motivates new optimizations and programming paradigms that would be either infeasible or inefficient with a device-granularity coherence strategy.
However, it also requires a methodology for efficiently selecting request types for all accesses in a program.

Section~\ref{sec:design-optimizations} describes two examples of coherence optimizations motivated by fine-grain coherence specialization (FCS). 
Section~\ref{sec:complexity} demonstrates how these optimizations can be implemented with minimal complexity overhead as long as a simple protocol baseline (e.g., Spandex) is selected.
Section~\ref{sec:design-req-select} presents a method for selecting request types for individual requests in a parallel heterogeneous workload.
Sections~\ref{sec:methodology}-\ref{sec:eval} demonstrate how these optimizations can improve performance for a range of workloads and access patterns.

\vspace{-1ex}
\subsection{New Coherence Request Types}
\label{sec:design-optimizations}

Two coherence optimizations that become feasible with FCS are forwarded update propagation and destination owner prediction.
We apply these optimizations on top of the baseline Spandex protocol (new request types shown in bold in Table~\ref{table:spandex_reqs}).
In designing these coherence optimizations, we require that new request types do not affect the consistency model (we use DRF) because using a different type of request type should only impact performance, never functionality.

\subsubsection{Forwarded Update Propagation (ReqWTfwd[+data]) }
The first coherence optimization adds a new type of update propagation method: forwarded write-through stores and RMWs (ReqWTfwd, ReqWTfwd+data).
When such a request arrives at the LLC for data that is owned by a different core, it will be forwarded to the current owner and its update operation will performed without affecting coherence state at the LLC or the current owner.
If the target data is not remotely owned, this request is treated the same as a normal ReqWT/ReqWT+data request.


\microhl{Write-through forwarding is useful when the consumer of a data update is expected to currently own it and the writer is not expected to exhibit more reuse than the consumer. 
This optimization motivates a new fine-grain specialized access pattern. 
By using ReqO+data for the consumer access (load or RMW) in a producer-consumer pair and ReqWTfwd[+data] for the producer access (store or RMW), 
the producer update will be forwarded to the owner and enable the performance-critical consumer access to exploit reuse that was previously infeasible.
Similarly, if the consumer and producer are atomic operations (e.g., an acquire and a release), they can use ReqO+data and ReqWTfwd+data, respectively.
This is an insight that has been exploited in the past, often in the context of update-based coherence. However, this optimization avoids the complex write atomicity challenges that plague the design of such protocols~\cite{SorinHill2011}.
This is because write-through forwarding only sends an update to a single exclusive owner core rather than all sharer cores, ensuring a single point of modification which serializes subsequent requests.}

\subsubsection{Destination owner prediction (ReqVo, ReqWTo)} 

\microhl{
The second optimization extends the interface for self-invalidated reads (ReqV) and forwarded write-through and RMW requests (ReqWTfwd[+data]) to allow a device to predict the current owner of target data and send a request directly to that core.
This effectively adds three new request types to the coherence interface:
ReqVo, ReqWTo, and ReqWTo+data. 
These represent variations of ReqV, ReqWT, and ReqWT+data requests for accesses that are good candidates for owner prediction. 
Once issued, these requests are handled the same as their root request types, except that they are sent directly to the expected owner core rather than the LLC (determined by an implementation-specific prediction mechanism).\footnote{A similar ownership prediction mechanism was considered as an extension to the DeNovo protocol~\cite{Komuravelli2015}, although the details were not discussed and the mechanism was not evaluated.} 
This is a simple extension in Spandex because, unlike read and write requests in most protocols, self-invalidated reads and forwarded write-through requests do not affect the LLC's coherence state. 
Thus, sending such requests directly to the current owner is functionally equivalent to looking up the current owner at the LLC.
Further, in the same way that a forwarded ReqV or ReqWTfwd request may fail and trigger a retry in Spandex, an incorrect owner prediction will simply trigger a retry with a non-forwarded (ReqWT or ReqO+data) request to the LLC.
Therefore, mispredicting an owner may hurt performance, but it does not affect protocol correctness.
}

If the underlying hardware supports owner prediction and the current owner can be accurately predicted, then using the owner prediction request types to directly forward read or write requests avoids the latency, traffic, and energy overheads of looking up the owner at the LLC.
Unlike write-through forwarding, this optimization requires an effective prediction mechanism to be worthwhile.
\iscahl{In this work we find that 
a simple hardware prediction mechanism is sufficient to improve performance for some common workloads. Our evaluation assumes a small prediction table is stored at each core which tracks the core ID of the most recent responder, indexed by PC and request type. 
This table is queried for every ReqVo and ReqWTo request and updated for every ReqVo and ReqWTo response.
As programming languages become more expressive and hardware-software codesign becomes more prevalent, software-provided ownership hints and finely-tuned hardware prediction techniques may be used to improve upon this strategy.}

\vspace{-1ex}
\subsection{Complexity Analysis}
\label{sec:complexity}

\sva{Unrelated to this section, but we need to reference and compare against CCIX and CXL clearly now that these specs are out.}

Coherence extensions (e.g., Section~\ref{sec:design}) are uncommon in practice primarily because they are difficult to design and validate when added to conventional protocols like MESI.\svac{reword slightly to just say they are difficult and focus on heterogeneous systems. implying they are uncommon in CPUs may open to attack so drop those words}
This is largely due to the rapidly growing state space that occurs when adding new transitions and request types to an already complex protocol~\cite{KomuravelliAdve2015}.

By building upon the simple and flexible DeNovo protocol, Spandex is easier to extend and optimize.
DeNovo relies on word granularity state and a DRF memory model to avoid transient blocking states~\cite{KomuravelliAdve2015}.
Mur$\phi$ is a model checking tool that takes a coherence protocol description and a set of system constraints and enumerates all reachable state vectors~\cite{Dill1996}.
The state vector includes the coherence state at a single controller, the state of data in each cache, and any pending or in-flight coherence requests.
Although not a perfect metric, we use the number of reachable state vectors in Mur$\phi$ as a proxy for protocol complexity.


We explore the Spandex protocols state \svac{vector?} space in Mur$\phi$ and compare it against a MESI protocol~\cite{MartinSorin2005} extended to implement a simplified version of the non-snoopable request types similar to those of the ARM CHI~\cite{CHISpec} interface (discussed more in Section~\ref{sec:related}). \svac{should we mention CCIX, CXL here - perhaps not? But better to motivate why we chose CHI and what's special about non-snoopable and how they relate to Spandex terminology}  
We use a flat cache organization instead of a hierarchical one~\cite{BeckmannGutierrez2015} because exhaustively exploring interactions between all devices in a multi-level protocol requires a prohibitively large state space.\svac{this hierarchical stuff comes from the blue - do we need more context or is it ok?}
Additionally, although Spandex supports variable request granularity, we do not explicitly model this in Mur$\phi$ and instead assume a fixed word granularity.
Spandex is expected to handle each word in a multi-word request individually based on its state. 
Therefore, we contend that adding support for coarse-grained or multi-word request types does not add any fundamental complexity to the protocol and would unfairly increase the state space in a Mur$\phi$ model.
Each of these protocols is then further extended with the write-through forwarding (ReqWTfwd) and owner prediction (ReqVo, ReqWTo) optimizations discussed in Section~\ref{sec:design-optimizations}.

\begin{figure}[tb!]
  \centering
  \includegraphics[scale=0.27]{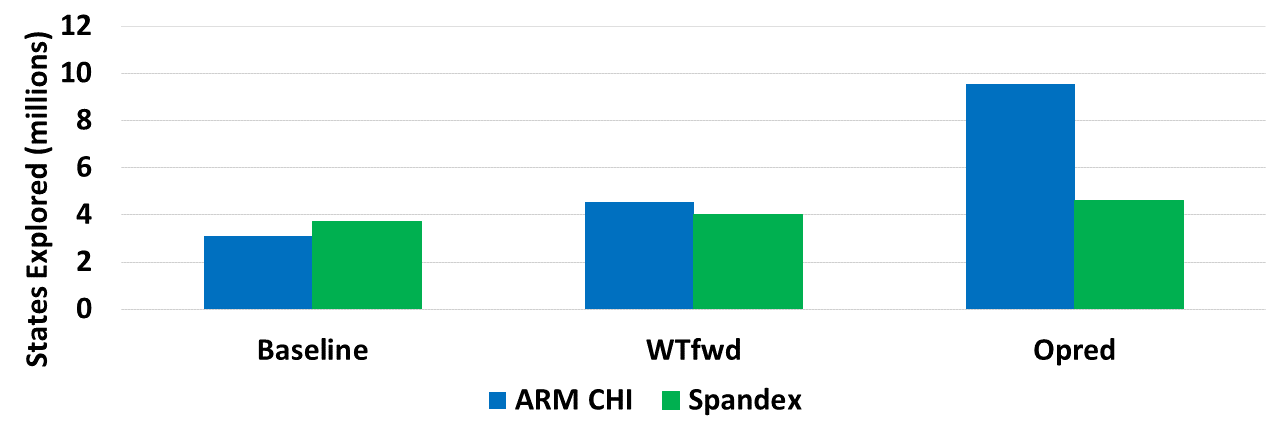}
  \vspace{-3ex}
  \caption{Comparing Spandex and ARM CHI's state spaces as optimizations are added.}
  \label{fig:mem-opt-complexity}
  \vspace{-1ex}
\end{figure}

Figure~\ref{fig:mem-opt-complexity} plots the number of states explored by Mur$\phi$ for each protocol.
Interestingly, the baseline Spandex protocol has slightly more states than the ARM CHI model because Spandex offers more flexibility. 
Where CHI has two types of load and store requests (snoopable and non-snoopable), Spandex allows the programmer to specify whether or not a store requires up-to-date data as well, and thus requires considering two more L1-initiated request types that may be pending at the L1 or LLC at any time.
This added flexibility allows a device to avoid false sharing and transient states that are difficult to avoid in fixed granularity protocols. 

However, as optimizations are added, the complexity of the MESI-based CHI coherence increases.
Compared with Spandex, which applies optimizations to non-blocking request types (ReqV, ReqWT), a MESI-based protocol must consider many corner case transient states for racy line granularity accesses. \sva{I don't think the above conveys what this fundamental complexity is well enough. Maybe I am confused with the previous paragraph.}
Thus, Mur$\phi$ explores 1.1x more states in ARM CHI with write-through forwarding added to both protocols, and 2.1x more states with owner prediction added.
This additional complexity stems from CHI being based on a line-granularity read-for-ownership protocol, and this prioritization is reflected in the state transitions.
Every cache miss causes a transition to a transient state that blocks subsequent requests until some satisfying set of messages is received to transition back to a stable state.
In contrast, Spandex builds upon DeNovo (which has no transient states) and avoids blocking states as much as possible. 
Thus, Spandex avoids much of the state space explosion that occurs when adding optimizations to a protocol with excessive transient states.
In fact, there is little increase in the number of Spandex states when write-through forwarding and owner prediction are added.

\vspace{-1ex}
\subsection{Request Type Selection: Overview}
\label{sec:design-req-select}

\begin{algorithm}[tb!]
\caption{\footnotesize Load request type selection (uses Algorithms~\ref{alg:owned_reuse_score}--~\ref{alg:owner_prediction_score}) 
}
\label{alg:load_selection}
\begin{algorithmic}
{\footnotesize
\IF{\textbf{OwnershipBeneficial}($X$)}
    \STATE $type$ = {\tt \textbf{ReqO+data}}
\ELSIF {\textbf{SharedStateBeneficial}($X$)}
    \STATE $type$ = {\tt \textbf{ReqS}}
\ELSIF{\textbf{OwnerPredBeneficial}($X$)}
    \STATE $type$ = {\tt \textbf{ReqVo}}
\ELSE
    \STATE $type$ = {\tt \textbf{ReqV}}
\ENDIF
}
\end{algorithmic}
\end{algorithm}

\begin{algorithm}[tb!]
\caption{\footnotesize Store request type selection (uses Algorithms~\ref{alg:owned_reuse_score} and~\ref{alg:owner_prediction_score})
}.
\begin{algorithmic}
\label{alg:store_selection}
{\footnotesize
\IF{\textbf{OwnershipBeneficial}($X$)}
        \STATE $type$ = {\tt \textbf{ReqO}}
\ELSIF{\textbf{OwnerPredBeneficial}($X$)}
    \STATE $type$ = {\tt \textbf{ReqWTo}}
\ELSE
    \STATE $type$ = {\tt \textbf{ReqWTfwd}}
\ENDIF
}
\end{algorithmic}
\end{algorithm}

\begin{algorithm}[tb!]
\caption{\footnotesize RMW request type selection  (uses Algorithms~\ref{alg:owned_reuse_score} and~\ref{alg:owner_prediction_score})
}
\begin{algorithmic}
\label{alg:rmw_selection}
{\footnotesize
\IF{\textbf{OwnershipBeneficial}($X$)}
        \STATE $type$ = {\tt \textbf{ReqO+data}}
\ELSIF{\textbf{OwnerPredBeneficial}($X$)}
    \STATE $type$ = {\tt \textbf{ReqWTo+data}}
\ELSE
    \STATE $type$ = {\tt \textbf{ReqWTfwd+data}}
\ENDIF
}
\end{algorithmic}
\end{algorithm}

\begin{algorithm}[tb!]
\caption{\footnotesize Request granularity selection for access $X$ of type $type$ (uses function definitions in Section~\ref{subsec:req_select_simple_func})
} 
\begin{algorithmic}
\label{alg:granularity_selection}
{\footnotesize
\IF{$type$ == {\tt \textbf{ReqV}}}
        \STATE $mask$ = \textbf{IntraSynchLoadReuse}($X$)
\ELSIF{$type$ == {\tt \textbf{ReqS}}}
        \STATE $mask$ = \textbf{FullBlockMask}($X$)
\ELSIF{$type$ == {\tt \textbf{ReqWT}} OR $type$ == {\tt \textbf{ReqWT+data}}}
        \STATE $mask$ = \textbf{RequestedWordsOnly}($X$)
\ELSIF{$type$ == {\tt \textbf{ReqO}} OR $type$ == {\tt \textbf{ReqO+data}}}
        \STATE $mask$ = \textbf{InterSynchStoreReuse}($X$)
        \IF{$mask$ != \textbf{RequestedWordsOnly}($X$)}
            \STATE $type$ = {\tt \textbf{ReqO+data}} 
        \ENDIF
\ENDIF
}
\end{algorithmic}
\end{algorithm}


\begin{algorithm}[tb!]
\caption{\footnotesize OwnershipBeneficial($X$) (uses function definitions in Section~\ref{subsec:req_select_simple_func})}
\label{alg:owned_reuse_score}
\begin{algorithmic}
{\footnotesize
\STATE {$Phase = 5, Xscore = 0, Y = X, Yprev = X$, $prevList=[X]$}


\algorithmicwhile $ $ $Y =$ \textbf{NextConflict}($Y$) \algorithmicdo
 \STATE \quad \algorithmicif $ $ \textbf{diffCores}($Yprev$, $Y$) \OR \textbf{SyncSep}($Yprev$, $Y$) \algorithmicthen
   \STATE \qquad $Phase = Phase - 1$
   \STATE \qquad \algorithmicif $ $ $Phase < 0$ \OR (\textbf{sameCore}($X$, $Y$) \AND  \NOT \textbf{reusePossible($X$, $Y$)}) \algorithmicthen $ $ break
     \STATE \qquad \algorithmicif $ $ \textbf{sameCore}($Y$, prevList) \algorithmicthen $ $ $Yval = 2* \mathbf{Criticality}(Y)$ 
     \STATE \qquad \algorithmicelse $ $ $Yval = 0.5*$ \textbf{Criticality}($Y$) 
     \STATE \qquad \algorithmicif $ $ \textbf{sameCore}($X$, $Y$) \algorithmicthen $ $ $Xscore = Xscore + Yval$
     \STATE \qquad \algorithmicelse $ $ $Xscore = Xscore - Yval$ ; $prevList.push(Y)$
   \STATE \quad $Yprev = Y$
\STATE \algorithmicif $ $ $Xscore > 0$ \algorithmicthen $ $ return $true$ \algorithmicelse $ $ return $false$
}
\end{algorithmic}
\end{algorithm}






\begin{algorithm}[tb!]
{\footnotesize
\caption{\footnotesize SharedStateBeneficial($X$) (uses function definitions in Section~\ref{subsec:req_select_simple_func})}
\label{alg:shared_reuse_score}
\begin{algorithmic}
\IF{\textbf{fromGPU}($X$)} 
\STATE return false
\ENDIF
\STATE {$Y = X, Yprev = X$}
\WHILE{$Y$ = \textbf{NextBlockConflict}($Y$)}
\IF{\textbf{diffCores}($Y$, $Yprev$) \OR \textbf{syncSep}($Y$, $Yprev$)}
    \IF{\textbf{isLoad}($Y$) \AND \textbf{sameCore}($X$, $Y$)}
        \STATE{return $true$}
    \ENDIF
    \IF{\textbf{isStore}($Y$) \AND \textbf{diffCores}($X$, $Y$)} 
        \STATE {return $false$}
    \ENDIF
    \STATE $Yprev = Y$
\ENDIF
\ENDWHILE
\STATE return $false$
\end{algorithmic}
}
\end{algorithm}

\begin{algorithm}[tb!]
\caption{\footnotesize OwnerPredBeneficial($X$) (uses function definitions in Section~\ref{subsec:req_select_simple_func})}
\label{alg:owner_prediction_score}
\begin{algorithmic}
{\footnotesize
\STATE {$Xscore = 0$}
\STATE {$Phase = 4, Xprev = $ \textbf{PrevConf}($X$)}
\STATE {$Y = X$} 
\WHILE{$Y = $ \textbf{PrevAcc}($Y$)} 
\STATE {$Yprev = $\textbf{PrevConf}($Y$)}
\IF{\textbf{sameCore}($X$, $Y$) \AND \textbf{sameType}($X$, $Y$)}
    \STATE $Phase = Phase - 1$ 
    \IF{$Phase < 0$}
    \STATE break
    \ENDIF
    \IF{\textbf{sameCore}($Yprev$, $Xprev$) }
        \STATE $Xscore = Xscore + 1$
    \ELSE
        \STATE $Xscore = Xscore - 1$
    \ENDIF
\ENDIF
\ENDWHILE
\IF{$Xscore > 0$}
    \STATE {return $true$}
\ELSE
    \STATE {return $false$}
\ENDIF
}
\end{algorithmic}
\end{algorithm}

Ideally, in a system with fine-grain coherence specialization, a compiler would automatically select the optimal request type for each instruction using static analysis. Developing such a compiler for arbitrary programs and heterogeneous systems is outside the scope of this work. 
Instead we develop (and implement) heuristic algorithms for determining appropriate request types for each access in a workload based on an execution trace. 

\asplos21hl{
Although the selection process is designed to be as hardware-agnostic as possible, we assume the following information is available to the selection mechanism:
\begin{itemize}
  \item Cache capacity (for determining long-term reuse potential) and cache block size (for spatial reuse potential).
  \item Whether or not the hardware supports the optimizations described in Section~\ref{sec:design-optimizations}.
  \item Whether each device is more likely to be sensitive to memory latency or bandwidth (i.e., whether it is a CPU or GPU).
\end{itemize}
If the above information is not available at selection time, multiple versions may be generated and a runtime can dynamically select from them when the information becomes available.
}
\asplos21hl{
Request selection considers a single sequentially consistent memory trace of a dynamic execution of a program and applies
Algorithms~\ref{alg:load_selection}-\ref{alg:rmw_selection} to each individual word-granularity memory access.\footnote{This methodology can also use byte-level access granularity, but this was not necessary for the workloads studied.}
The ordering of accesses and the issuing devices are therefore known and leveraged by the selection algorithm.\footnote{"Device" is used to refer to any computing element which has a single local private cache, including individual CPU and GPU cores.}
For instructions that access multiple words, request selection treats each word as a separate access, and these word accesses vote on a request type for the full dynamic access.
}

Algorithm~\ref{alg:load_selection} chooses between using a {\tt \textbf{ReqV}}, {\tt \textbf{ReqS}}, or {\tt \textbf{ReqO+data}} for load accesses based on whether obtaining ownership is expected to improve global cache reuse, obtaining shared state will improve global cache reuse, or neither will improve reuse (motivating valid state).

Algorithm~\ref{alg:store_selection} chooses between using a {\tt \textbf{ReqO}} or {\tt \textbf{ReqWTfwd}} for store accesses based on whether obtaining ownership is expected to increase global cache reuse.

Algorithm~\ref{alg:rmw_selection} chooses {\tt \textbf{ReqO+data}} or {\tt \textbf{ReqWTfwd+data}} for an atomic RMW access based on whether obtaining ownership is expected to increase global cache reuse.


Finally, Algorithm~\ref{alg:granularity_selection} determines whether an access should request more words in the cache line than are requested by the original instruction if this is expected to increase cache reuse.

All of the above algorithms use simple functions (defined in Section~\ref{subsec:req_select_simple_func}) and heuristic algorithms (defined in Algorithms~\ref{alg:owned_reuse_score}-\ref{alg:owner_prediction_score}, explained in Section~\ref{subsec:req_select_heuristics}) for improved clarity.

\vspace{-1ex}
\subsection{Basic Request Selection Functions}
\label{subsec:req_select_simple_func}

Some function definitions are straightforward.
\textbf{isLoad}($X$),
\textbf{isStore}($X$),
\textbf{fromCPU}($X$),
and \textbf{fromGPU}($X$)
return true if $X$ is a load, $X$ is a store, $X$ is issued from a CPU, or $X$ is issued from a GPU, respectively.
\textbf{sameCore}($X$, $Y$)
and \textbf{diffCores}($X$, $Y$)
return true if $X$ and $Y$ are issued from the same core or different cores, respectively. 
\textbf{sameInst}($X$, $Y$)
returns true if $X$ and $Y$ are dynamic instances of the same static instruction.

\noindent 
\asplos21hl{
\textbf{ReusePossible}($X$, $Y$) 
approximates whether the data accessed by $X$ will still be in the cache when access $Y$ is executed, and it returns true only if the reuse distance (defined here as the number of unique bytes accessed between $X$ and $Y$ from the issuing core in the dynamic trace)
is less than $N$, where $N$ is set to be 75\% of the cache capacity. 
}

\noindent 
\asplos21hl{
\textbf{NextConflict}($X$) returns the next access to
the same address as $X$ following $X$ in the dynamic trace.
It is used to determine whether subsequent accesses to a given variable will be issued from the same core (potentially enabling cache reuse) or from a different core (potentially interfering with reuse).
\textbf{NextBlockConflict}($X$) returns the next access to
any address in the cache block accessed by $X$ following $X$ in the 
trace.
}

\noindent 
\asplos21hl{
\textbf{PrevAcc}($X$) returns the most recent access preceding $X$ in the dynamic trace.
\textbf{PrevConf}($X$) returns the most recent access to
the same address as $X$ preceding $X$ in the 
trace.
}

\noindent 
\asplos21hl{
\textbf{SyncSep}($X$, $Y$) is true if $X$ and $Y$ are issued from the same core and there exists a synchronization operation $S$ between $X$ and $Y$ in program order such that either 1) $X$ or $Y$ is an atomic access, 2) $X$ is a load and $S$ is an acquire operation (e.g., the start of a new kernel), or 3) $X$ is a store and $S$ is a release operation (e.g., the completion of a kernel). 
Since synchronization accesses can invalidate and flush the cache of Valid data and atomic accesses can only hit on owned state,\svac{not true in general; e.g., for mesi} this helps determine whether obtaining Owned or Shared permission is required to exploit available reuse. 
}

\noindent 
\textbf{Criticality}($X$) returns a value that represents the estimated 
\noindent performance criticality of $X$.
We use a very simple heuristic for our evaluation of CPU-GPU workloads.\svac{first time mentioning that we are doing CPU GPU. do we want to say something more general also?}
Loads and atomic RMW accesses without release semantics\footnote{Synchronization accesses with release semeantics generally don't use the return value and are therefore less critical.} tend to be more performance critical than stores and release atomics, and CPU accesses tend to be more sensitive to access latency than GPU loads.
Therefore, we assign CPU loads and non-release atomic RMW accesses a criticality weight of 6 and GPU loads and non-release atomic RMW accesses a criticality weight of 2.
All other access types have a criticality weight of 1.



\noindent 
\textbf{IntraSynchLoadReuse}($X$) 
returns a word mask 
based on whether requesting valid state for additional words in a cache block is expected to improve reuse.
For each word 
in the cache block accessed by $X$, this function sets the corresponding mask bit if there is a subsequent load $Y$ to that word address which will not be invalidated before it is accessed (i.e., is not separated by a synchronization action).\svac{again, this is only for GPUs. reword}
Specifically, this requires that $Y$ satisfies the following properties: 1) \textbf{sameCore}($X$, $Y$), 2)  \textbf{reusePossible}($X$, $Y$), and 3) !\textbf{syncSep}($X$, $Y$).

\textbf{InterSynchStoreReuse}($X$) returns a word mask 
based on whether requesting ownership for additional words in a cache block is expected to improve reuse.
For each word 
in the cache block accessed by $X$, this function sets the corresponding mask bit if there is a subsequent store 
$Y$ to that word address which cannot be coalesced in a write combining buffer (i.e., is separated by a synchronization action).
Specifically, this requires that $Y$ satisfies the properties: 1) \textbf{sameCore}($X$, $Y$), 2)  \textbf{reusePossible}($X$, $Y$), and 3) \textbf{syncSep}($X$, $Y$).
\vspace{-1ex}
\subsection{Heuristic Request Selection Algorithms}
\label{subsec:req_select_heuristics}

\noindent
\textbf{OwnershipBeneficial}($X$) is a heuristic that returns true if ownership for $X$ is likely to improve overall performance 
and is defined in Algorithm~\ref{alg:owned_reuse_score}.
At a high level, ownership is beneficial for an access if the issuing core is the most common or most critical requestor in subsequent accesses to the same variable.

\asplos21hl{
To approximate whether this is the case, each access $Y$ following $X$ and to the same address as $X$ in the SC execution is assigned a value based on whether $X$ obtaining ownership will help or hurt performance for $Y$, 
and improved performance is expected if the sum of these is greater than zero.
An access $Y$ is ignored if the previously considered access in the SC memory trace was to the same core and the two are not sync-separated (this is because the lack of intervening synchronization means that reuse is possible for this access regardless of ownership state).
Otherwise, $Y$ is assigned a value based on whether the issuing core has accessed the data recently (indicating reuse potential for $Y$) and 
the criticality of $Y$ (it is more important to exploit reuse for reads and CPU accesses).
If $Y$ is issued from the same core as $X$, then ownership would help and its value is added to the running sum; if $Y$ is issued from a different core, then ownership for $X$ would hurt performance for $Y$ and its value is subtracted from the running sum.
This process ends once a fixed number of subsequent \svac{(sync-separated)} $Y$ accesses are evaluated (5 in our algorithm).
}

\noindent 
\textbf{SharedStateBeneficial}($X$) is a heuristic that returns true if obtaining shared state for $X$ would improve performance based on expected reuse effects, and is defined in Algorithm~\ref{alg:shared_reuse_score}. \svac{The numbers for the algorithms are printing wrong}
We assume that obtaining shared state is beneficial if the issuing core is a CPU \svac{we should generalize this beyond CPU to a latency-critical compute unit? also see new comment in the algm} (it is more sensitive to cache miss overheads than sharer tracking overheads) and it can lead to at least one future cache hit.
This is true if there is at least one load to the same cache block from the same core, it is sync-separated with $X$ (otherwise reuse is possible with or without shared state), and there is no intervening remote store to that cache block (this would invalidate the shared state).

\noindent 
\textbf{OwnerPredBeneficial}($X$) is a heuristic that returns true if using owner prediction is likely to be successful for the given prediction mechanism based on execution history,
and is defined in Algorithm~\ref{alg:owner_prediction_score}.
In the evaluated system, each device \svac{is this a device (which to me means CPU or GPU) or is it a compute unit (which to me means core or SM)} uses a simple prediction mechanism which predicts an owner based on recent responding devices for requests of the same type as $X$.
We assume ownership prediction will be beneficial if the most recent prior accesses from the issuing core had an immediate predecessor access from the same \svac{same or different -- I am having trouble with this sentence}core as the immediate predecessor of $X$.
\asplos21hl{
To approximate this, we work backwards from $X$, considering previous accesses to the same address, from the same core, and of the same type (load, store, or RMW) as $X$.
Similar to the ownership benefit heuristic, a score is calculated to determine potential benefit.
If the immediate predecessor of an access $Y$ \svac{Y has to be in the same core as X as I think - this text needs tweaking}in the SC execution is issued by a remote core and that core is the same core that issues the immediate predecessor to $X$, then accurate prediction is more likely and that access's value is added to the score. If not, then the value is subtracted from the score.
Again, magnitude depends on proximity to $X$ (more recent accesses affect the predictor the more), and evaluation stops after a fixed number of prior accesses (5 in our algorithm) have been considered.
}

\vspace{-2ex}
\subsection{Incomplete Request Type Support}
\label{sec:design-incomplete_support}
\vspace{-1ex}

\asplos21hl{
Some systems may not support all of the optimized request types described in the above algorithms.
If owner prediction is not supported, then OwnerPredBeneficial always returns false. 
If write-through forwarding is not supported by a core, then producer-consumer forwarding is no longer possible.
Thus, Criticality($X$) should return the same value for loads and stores (consumers should not be preferred for ownership), and selection should then proceed as described above.
After selection has completed, forwarded requests must be converted to supported request types.
Stores assigned ReqWTfwd should use ReqWT instead.
Atomic RMW operations assigned ReqWTfwd+data should use ReqO+data if both the prior and subsequent accesses in the dynamic trace use ownership, and ReqWT+data otherwise.\footnote{Stores and RMWs differ because, unlike a ReqWT, a ReqWT+data can cause an ownership revoke and excessive data transfer, so obtaining ownership may be preferred even if no reuse is possible.}
If word granularity state is not supported by a core (state is stored at cache line granularity), then that core can use full block mask for each request, 
and any ReqO requests must be converted to ReqO+data requests. 
}

\vspace{-1ex}
\section{Target Workloads}
\label{sec:workloads}
\vspace{-1ex}

Since current heterogeneous systems are not designed for fine-grain coherence specialization, there are no standard benchmarks.
Instead, we identify two sets of workloads which exhibit more request diversity than is common in conventional heterogeneous applications, which we refer to as microbenchmarks and applications.
We use the request selection algorithms from Section~\ref{sec:design-req-select} for all benchmarks.

\vspace{-1ex}
\subsection{Microbenchmarks}
\label{subsec:methodology-ubmks}
\vspace{-1ex}

\begin{figure*} 
  \centering
  \subfloat[FlexV/S]{ 
    \label{fig:flex_V_S}%
    \includegraphics[scale=0.2]{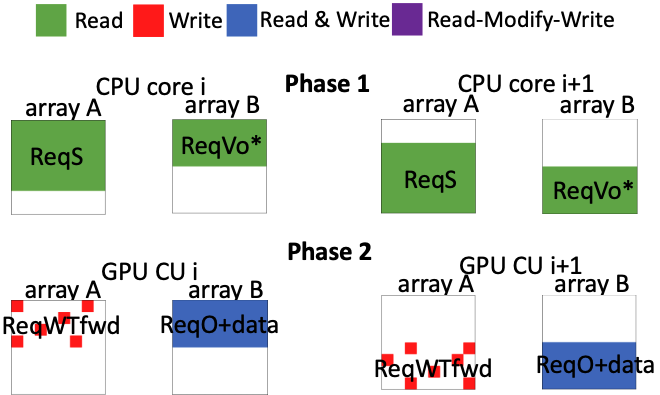}%
  }
  \hspace{8pt}%
  \subfloat[FlexO/WT]{
    \label{fig:flex_O_WT}%
    \includegraphics[scale=0.2]{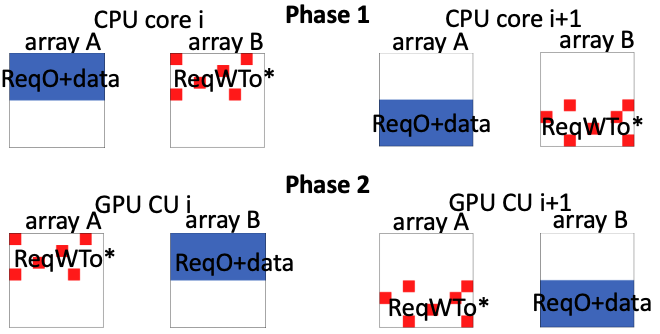}%
  }
  \hspace{8pt}
  \subfloat[FlexOa/WTa]{
    \label{fig:flex_Oa_WTa}%
    \includegraphics[scale=0.35]{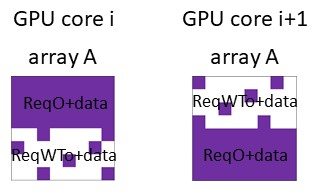}%
  }
  \hspace{8pt}%
  \subfloat[Prod-Cons]{
    \label{fig:indirection}%
    \includegraphics[scale=0.2]{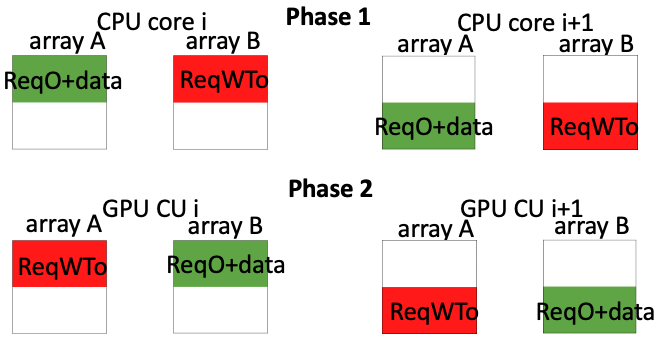}%
  }
  \caption{Access patterns and coherence request types used by fine-grain coherence specialization for microbenchmarks. All accesses except those marked with * are to the same partition in each subsequent iteration of the algorithm.} 
  \svac{One way to make space here is to put these in one column. That way, the white space in figure c can be saved and we should use the same font size for that one. Reduce the spacing between the arrays under a single core further so different cores are easier to identify. I also think that we could play with width and height of the different arrays to make more space and perhaps even fit everything in one row. There is a lot of white space here that is not needed; e.g., between the two arrays. We can also put vertical lines between different benchmarks to separate them and pack them tighter in a row. The legend is taking up too much space too (that will go away if we do it all in one column; if we do it in one row, then the legend could be under the atomics benchmark)}
  \label{fig:ubmk_figs}%
  \vspace{-3ex}
\end{figure*}

We use four microbenchmarks to isolate the precise benefits of fine-grain coherence specialization.
Each microbenchmark has multiple phases in which CPU and/or GPU cores concurrently access shared data, with cores synchronizing between each execution phase.
These phases are iteratively executed.


\subsubsection{FlexV/S} 
This microbenchmark illustrates the benefits of dynamically choosing between self-invalidated reads (ReqV) and writer-invalidated reads (ReqS).
As in Figure~\ref{fig:flex_V_S}, 
CPU cores densely read array A, and these accesses exhibit both sharing (multiple CPUs read the same data) and sync-separated reuse (each core accesses the same partition in successive CPU phases). 
CPU cores also densely read array B with no sharing and no reuse (each core accesses different partitions in successive CPU phases). 
GPU cores sparsely write array A with no reuse, and densely read and write array B with high reuse. 
This workload is representative of any workload which must read from two data structures: one which exhibits inter-phase reuse and sharing 
and one which exhibits no reuse. 
For example, a weight matrix in a DNN layer will exhibit reuse and sharing because the same weight values will be accessed by all cores processing the layer in every iteration,  
while the feature inputs will read different data in each iteration (e.g., if a queue or double-buffering is used for inputs).

CPU reads to array A exhibit reuse across phases and are unlikely to be separated by a remote write, but multiple CPUs access the same data concurrently. Thus, shared state is beneficial and ownership is not beneficial, resulting in the use of ReqS.
CPU reads to array B exhibit no reuse, but the remote core is likely to be predicted based on the last responder (the partition of B accessed by a CPU core in a given phase will have been produced by the same GPU core). Therefore, ReqVo is used for these accesses.
GPU stores to A exhibit low inter-phase reuse, so ReqWTfwd is used. 
GPU stores to B exhibit high inter-phase reuse and are not accessed more frequently by any remote core, so ownership is beneficial and ReqO+data is used for these accesses.

\subsubsection{FlexO/WT}
This microbenchmark illustrates the benefits of dynamically choosing between ownership-based data stores (ReqO) and write-through data stores (ReqWT).
As illustrated in Figure~\ref{fig:flex_O_WT},
CPU cores densely read and write the same partition of array A and sparsely write array B. Then GPU cores densely read and write array B and sparsely write array A. 
The dense CPU and GPU accesses are to the same partition in successive iterations, while the sparse CPU and GPU accesses are to different partitions in successive iterations.
This represents workloads which update different data structures with different reuse levels in a non-racy manner (e.g., machine learning models with both sparse and dense layers like the Deep Learning Recommendation Model~\cite{NaumovMudigere2019}).

The dense CPU and GPU accesses exhibit sync-separated reuse, and ownership is beneficial because each element is unlikely to be accessed by a remote core between the reuse accesses.
Therefore, these accesses request ownership (ReqO+data for reads, ReqO for writes). 
The sparse CPU and GPU writes exhibit low reuse potential, 
and owner prediction is feasible because the same remote core will be the most recent owner of all data accessed by a core in a given phase.
Therefore, ReqWTo is used for these accesses.

\subsubsection{FlexOa/WTa}
This microbenchmark illustrates the benefits of dynamically choosing between ownership-based atomic accesses (ReqO+data) and write-through atomic accesses (ReqWT+data).
As illustrated in Figure~\ref{fig:flex_Oa_WTa}, 
a single phase and single core type is sufficient for this purpose since atomics enable racy accesses and communication.\footnote{GPU-only is studied, but a CPU-only version would exhibit the same request types and tradeoffs.}
In each iteration, each core densely reads and writes its local partition and sparsely reads and writes a remote partition. 
This is representative of any workload which distributes a data structure across multiple cores and allows each core to sparsely access remote partitions concurrently with remote cores (e.g., racy updates to a hybrid dense+sparse machine learning model, or push-based graph workloads with densely connected components).\svac{give citations to these things if possible.}

CPU accesses to A's local partition exhibit high sync-separated reuse and intervening remote accesses are rare, so ownership is beneficial and these RMWs use ReqO+data requests. 
Sparse accesses to remote partitions exhibit low sync-separated reuse, and the remote owner can be predicted within a phase. 
Therefore, these accesses use ReqWTo+data.

\subsubsection{Prod-Cons}
This microbenchmark illustrates the benefits of write-through forwarding and owner prediction.
In each iteration, CPUs read a partition of array A and write a partition of array B.
GPUs then read a partition of array B and write a partition of array A.
This is representative of streaming workloads, data transform kernels, or applications that transfer input and output data to and from a specialized accelerator (e.g., the EP algorithm discussed in Section~\ref{subsubsec:method_ep}).

Each core reads the same partition and writes the same partition in each iteration, so there is sync-separated reuse for all accesses.
When write-through forwarding is supported, the criticality and reuse of consumer reads will outweigh the locality benefits for producer writes such that ReqO+data will be used for reads, and ReqWTo will be used for writes (owner prediction is feasible due to the regular access pattern).
If write-through forwarding is not supported, then reads are not rated more highly in the evaluation of ownership benefit, so ReqVo is used for reads 
and ReqO is used for writes.

\vspace{-1ex}
\subsection{Applications}
\label{subsec:methodology-pipe-workloads}
\vspace{-1ex}

In addition to the above microbenchmarks, we evaluate four full applications which can benefit from fine-grain coherence specialization: a fully connected neural network (FCNN), a convolutional neural network (LeNet), a recurrent neural network (LSTM), and an evolutionary programming algorithm (EP).
FCNN, LeNet, and LSTM are GPU-only neural network workloads, all of which assume a batch size of 1.\footnote{Larger 
batch sizes can improve 
cache reuse, but this comes at the cost of end-to-end latency, which is often unacceptable for latency sensitive inference workloads running on resource-constrained systems~\cite{FowersOvtcharov2018,MohammadiAlFuqaha2018}.} 
EP is a CPU+GPU application which demonstrates how fine-grain coherence specialization can benefit applications that offload only some tasks to a specialized accelerator (the GPU). 

\subsubsection{Fully Connected Neural Network (FCNN)}

Fully connected networks have proven useful for processing inputs that are spatially and temporally unrelated such as medical~\cite{SaponAkmal2011}, real estate~\cite{ChopraThampy2007}, and recommendations~\cite{CovingtonAdams2016}.


We compare data parallel and pipelined implementations of a neural network consisting of 5 uniform, fully connected layers. 
In a data parallel implementation, each device independently processes all layers for a distinct subset of the input features, avoiding the need for inter-device communication but preventing reuse for weight data. 
In a pipelined implementation, each device executes a single stage of the computation for every input.
Although a pipelined implementation requires inter-device communication along producer-consumer edges, it is able to exploit reuse in weight accesses because each device only needs to load a single weight matrix for all inputs. 

Both versions use 5 GPU CUs to process 60 independent input features.
In each iteration, each layer multiplies an input vector with a weight matrix and performs a ReLU operation.
In the pipelined implementation, inputs and outputs are double-buffered to enable concurrent pipelined execution, and atomics are used to synchronize between adjacent layers.
Specialized coherence is enabled only for the pipelined implementation (since data parallel does not benefit from specialization).
Ownership is beneficial for the weight matrix and input buffer loads (which use ReqO+data), while output buffers stores use write-through forwarding and owner prediction (ReqWTo). Atomic release accesses use write-through forwarding (ReqWTo+data), while acquire accesses use owned atomics (ReqO+data). This constitutes producer-consumer forwarding for the synchronizaing atomics.


\subsubsection{Convolutional Neural Network (LeNet)}
To classify spatially related data such as images, some of the most common neural networks in deep learning are primarily composed of convolutional layers with a few fully connected layers at the end. 
One such example is LeNet, which has been used to determine the numerical values of handwritten digits~\cite{LecunJackel1995}. 

\hl{
LeNet consists of two convolutional layers, two maxpool layers, and two fully connected layers~\cite{Lecun2015}.
Again, we evaluate both a data parallel version and a pipelined parallel version.
Both pipelined and data parallel versions use 10 GPU CUs to process 80 independent input features.
The data parallel implementation maps layers to cores in the same way as FCNN.
However, the variable work required at each layer in LeNet complicates the pipelined implementation; load imbalance between nodes in the pipeline can lead to performance bottlenecks, since inputs can only be processed as fast as the slowest layer.
To improve load balance, long-running layers have been split among multiple cores where possible.
Like FCNN, loads to the weight matrices benefit from ownership and use ReqO+data when fine-grain coherence specialization is enabled, and atomic accesses use producer-consumer forwarding. 
However, feature data reads do not benefit from ownership, since the feature data produced at each  layer is concurrently read by multiple consumers.
Therefore, producer-consumer forwarding is not possible for LeNet features.
}

\subsubsection{Recurrent Neural Network (LSTM)} 
Recurrent neural networks can be used to classify sequentially associated data series such as audio~\cite{AmodeiAnanthanarayanan2016}, video~\cite{YaoTorabi2015}, DNA~\cite{ParkMin2017}, ECG data~\cite{ChauhanVig2015}, and aircraft sensor data~\cite{NanduriSherry2016}.
We evaluate a long short-term memory (LSTM) RNN model proposed for network anomaly detection~\cite{RadfordApolonio2018}. The model uses two stacked LSTM layers, a single dense layer activation, and a single fully connected softmax output layer, with a ten-token input sequence to predict the subsequent (eleventh) token. 

Since the inputs have serial dependencies, the only way parallelize a single input sequence is with pipelined parallelism. 
Each LSTM layer is composed of 50 hidden cells, 50 memory cells, a 50-entry input vector, a 50-entry output vector, and four 50x100 entry weight matrices. 
The layers are distributed among CUs; each LSTM layer is assigned to a CU, while the dense layer and softmax layer are fused and assigned to one CU.
As with pipelined FCNN and LeNet, double-buffering and atomic accesses are used to synchronize between adjacent layers. 
With fine-grain coherence specialization, loads to the weight matrices and input buffers use ReqO+data, stores to output buffers use ReqWTo. 
Synchronizing atomic accesses use producer-consumer forwarding. 

\subsubsection{Evolutionary Programming (EP)}
\label{subsubsec:method_ep}

Evolutionary Programming mimics the natural selection process in a simulated population to minimize a cost function. 
It consists of repeated execution of five consecutive stages: reproduction, evaluation, selection, crossover, and mutation~\cite{sun2016heteromark}. 
We use parameters used by Capraro et al. for radar waveform selection~\cite{Capraro08} where population is 330 and mutation and crossover rates are set to 0.5. 
For the mutation stage, we perform 
a centre inverse mutation (CIM) where each chromosome is divided into two sections and all genes in each section are copied and then inversely placed in the same section of a child~\cite{Abdoun12}. For task-partitioning, we use HeteroMark's implementation which offloads the \textit{evaluation} and \textit{mutation} stages to the GPU, and executes the remaining stages on the CPU~\cite{sun2016heteromark}.
With fine-grain coherence specialization, we find that much of the data transferred between CPU and GPU experiences high reuse on both devices.
Since CPU reuse is weighted higher than GPU reuse in the selection algorithms (Section~\ref{sec:design-req-select}), our specialized coherence implementation obtains ownership for CPU reads (ReqO+data) and forwards GPU writes to the CPU to improve CPU reuse (ReqWTo), at the cost of GPU reuse.

\vspace{-1ex}
\section{Methodology}
\label{sec:methodology}



\vspace{-1ex}
\subsection{Coherence Configurations}
\label{subsec:methodology-staticConfigs}
\svac{make sure we say early on that we are doing CPU and GPU because [something like -] no implementations with accelerators etc. or it is too hard to do that right now}
We compare seven different configurations of device cache architectures.
The first four configurations use a static request type selection at each device cache corresponding to the MESI, DeNovo, or GPU coherence protocols. 
All L1 caches interface directly with the Spandex LLC. 

\noindent
\textbf{SMG}: Static - MESI CPU caches, GPU coh. GPU caches

\noindent
\textbf{SMD}: Static - MESI CPU caches, DeNovo GPU caches

\noindent
\textbf{SDG}: Static - DeNovo CPU caches, GPU coh. GPU caches

\noindent
\textbf{SDD}: Static - DeNovo CPU caches, DeNovo GPU caches

The static protocol configurations are chosen based on the suitability of each protocol for CPU and GPU memory demands.
The final three configurations use fine-grain coherence specialization with neither write-through forwarding and owner prediction (\textbf{FCS}), with only write forwarding (\textbf{FCS+fwd}), and both (\textbf{FCS+pred}).
We  choose the memory request type according to Section~\ref{sec:design-req-select}, \svac{again, reword for Sam's tool} and owner prediction is based on a per-core prediction table which stores the last responder's ID for each Spandex request type.

\vspace{-1ex}
\subsection{Simulation Environment and Parameters}
\label{subsec:methodology-sim}
\vspace{-1ex}
We execute the above workloads and configurations using an integrated CPU-GPU architecture simulator.
GPGPU-Sim~\cite{BakhodaYuan2009} models the GPU CUs \iscarhl{(we use an architecture similar to NVIDIA GTX480)}.
\iscarm{GPU kernel launch latency is not modeled, apart from the cache flush latency at kernel boundaries.} \svac{need to say that this makes our results conservative if that is the case. In any case, need to be clear about the impact of this or it raises a red flag that we bias in our favor}
Simics~\cite{MagnussonChristensson2002} models the CPU cores, Wisconsin GEMS~\cite{MartinSorin2005} models the memory timing, and Garnet~\cite{AgarwalKrishna2009} models the network.
There are 16 CPU cores and 16 GPU CUs, connected through a 4x4 mesh (a CPU core and GPU CU at each node).
Each CPU core and GPU CU has its own L1 cache, and these caches interface directly with a Spandex LLC. 
\iscarm{All cores use a write buffer, which coalesces pending writes and is flushed at synchronization points. If coalesced writes have different request types, the highest priority request type is used, following this prioritization: ReqWT $>$ ReqO $>$ ReqWTfwd $>$ ReqWTo.} 
Table~\ref{table:sim} lists the detailed simulation parameters.

\svac{Sam's tool -- a separate subsection to highlight it will be good, especially if can run the apps through.}

\begin{table}[tb!]
{\footnotesize
  \centering
  \begin{tabular}{|c|c|}
    \hline
    \multicolumn{2}{|c|}{\textbf{CPU/GPU Parameters}} \\ \hline
    Frequency & {2 GHz / 700 MHz} \\ \hline
    Cores & {16 / 16} \\ \hline
    \multicolumn{2}{|c|}{\textbf{Memory Hierarchy Parameters}} \\ \hline
    L1 Size (32-way assoc.) & 128 KB \\ \hline
    L2 Size (16 banks, 32-way assoc.) & 8MB \\ \hline
    L1 MSHRs, Write Buffer, and Store Buffer Sizes & {128 entries each} \\ \hline
    L1 hit latency in cycles & {1} \\ \hline
    Remote L1 hit latency in cycles& {135$-$183} \\ \hline
    L2 hit latency in cycles & {129$-$161}\\ \hline
    Memory latency in cycles& 297$-$361\\ \hline
  \end{tabular}
  \vspace{-1ex}
  \caption{Simulated heterogeneous system parameters.}
  \label{table:sim}
  \vspace{-3ex}
}
\end{table}


\vspace{-1ex}
\section{Evaluation}
\label{sec:eval}


\begin{figure}
  \centering
  \begin{tabular}{@{}c@{}}
    \includegraphics[width=0.45\textwidth]{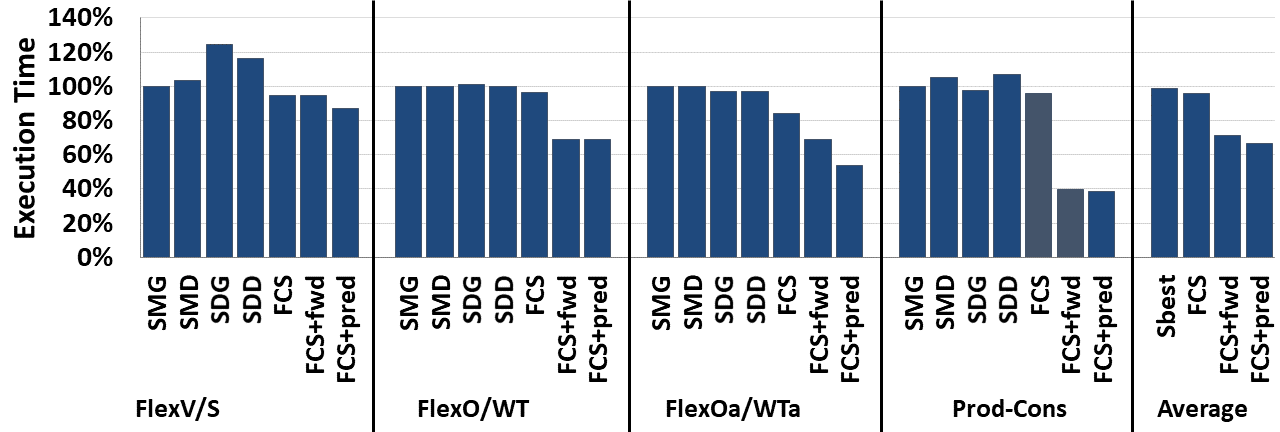} \\[\abovecaptionskip]
    \vspace{-2ex}
    \small (a) Execution time normalized to SMG.
    \label{fig:mpe-ubmks-cycles}
  \end{tabular}
  
  \vspace{\floatsep}
  
  \begin{tabular}{@{}c@{}}
    \includegraphics[width=0.45\textwidth]{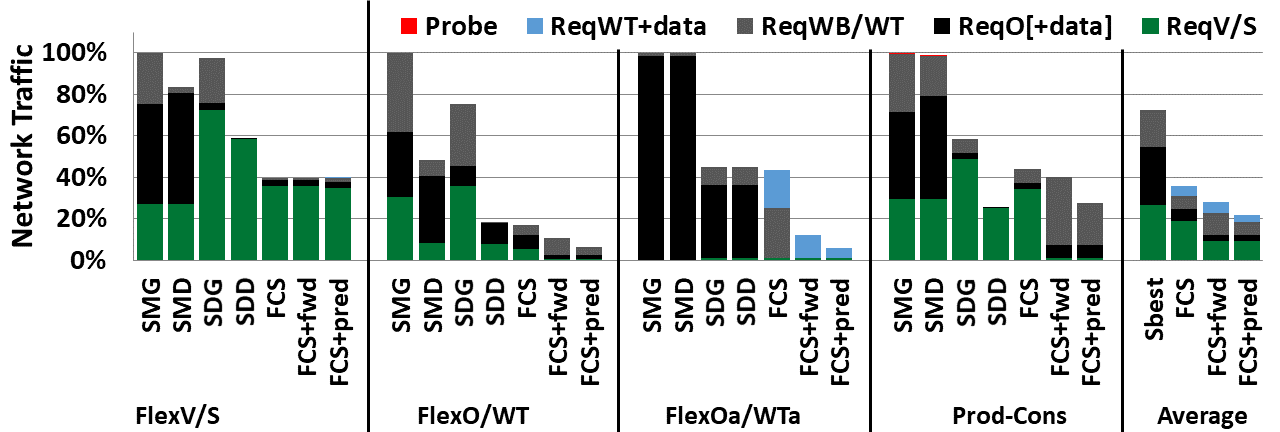} \\[\abovecaptionskip]
    \vspace{-2ex}
    \small (b) Network traffic normalized to SMG.
    \label{fig:mpe-ubmks-nw}
  \end{tabular}
  \vspace{1ex} 
  \caption{\fontsize{9pt}{11pt} Microbenchmark's execution time and network traffic.}
  \label{fig:mpe-ubmks}
  \vspace{-3ex}
\end{figure}


\vspace{-1ex}
\subsection{Microbenchmark Results}
\label{subsec:eval-ubmks}

Figure~\ref{fig:mpe-ubmks} shows the execution time and network traffic for the microbenchmarks. 
Sbest represents the best performing (lowest execution time) static coherence configurations for each microbenchmark averaged over all microbenchmarks.


\noindent
\textbf{FlexV/S}: 
Obtaining Shared state for the read accesses to array A improves CPU cache hit rates, reducing latency and traffic for these CPU read accesses.
Obtaining Shared state for the CPU read accesses to array B with no inter-kernel reuse is wasteful and can require revoking ownership from a remote GPU core, preventing reuse there. 

FCS exploits inter-phase CPU reuse for array A while also avoiding wasteful Sharer overheads (e.g., revoking ownership from the GPU) for array B.
This improved reuse reduces network traffic by 60\% relative to the fastest static configuration (SMG), although execution time is only reduced by 5\% due to the GPU's ability to tolerate memory latency.
FlexV/S does not benefit from write-through forwarding because written data is never remotely owned, but owner prediction is helpful for CPU reads to A, reducing execution time by a further 8\%.


\noindent
\textbf{FlexO/WT}: 
FCS reduces execution time and network traffic by 7\% each relative to the best static configuration (SDD) by using ReqWT for sparse writes. 
This revokes ownership from dense owners, but subsequent dense accesses incur only 2-hop misses 
(versus 3-hops if the sparse access obtained ownership).
With FCS+fwd, execution time and network traffic are reduced a further 13\% and 38\%, respectively, because sparse writes do not revoke ownership from dense owners. 
Ownership prediction avoids LLC lookups for forwarded stores (ReqWTo), 
reducing network traffic a further 19\%.

\noindent
\textbf{FlexOa/WTa}: 
As with FlexO/WT, FCS 
reduces execution time by 14\% and network traffic by 3\% relative to the fastest static configuration (SDD) because the write-through sparse accesses reduce indirection for future dense accesses.
When write-through forwarding is enabled, sparse atomics are forwarded to the core that owns the target partition, improving cache hit rates for dense accesses and reducing execution time and network traffic by a further 18\% and 72\%, respectively.
Owner prediction further reduces indirection for sparse accesses, reducing execution time and network traffic by a further 22\% and 50\%, respectively.

\noindent
\textbf{Prod-Cons}: 
FCS is unable to improve performance beyond SDG for Prod-Cons, but with forwarding enabled (FCS+fwd) it can avoid revoking ownership from latency-sensitive consumers. 
By forwarding writes to the consumer, FCS+fwd moves the overhead of data movement from the reader to the writer and reduces
execution time by 58\% relative to the fastest static configuration (SDG).
However, since forwarded write-through data must travel two hops rather than one, FCS+fwd increases network traffic by 35\% relative to SDD. 
This overhead can be avoided by owner prediction, 
and FCS+pred reduces network traffic by 19\% relative to SDD.



\vspace{-1ex}
 \subsection{Application Results}
\label{subsec:eval-pipe}
\begin{figure}[tb!]
  \centering
  \begin{tabular}{@{}c@{}}
    \includegraphics[width=0.45\textwidth]{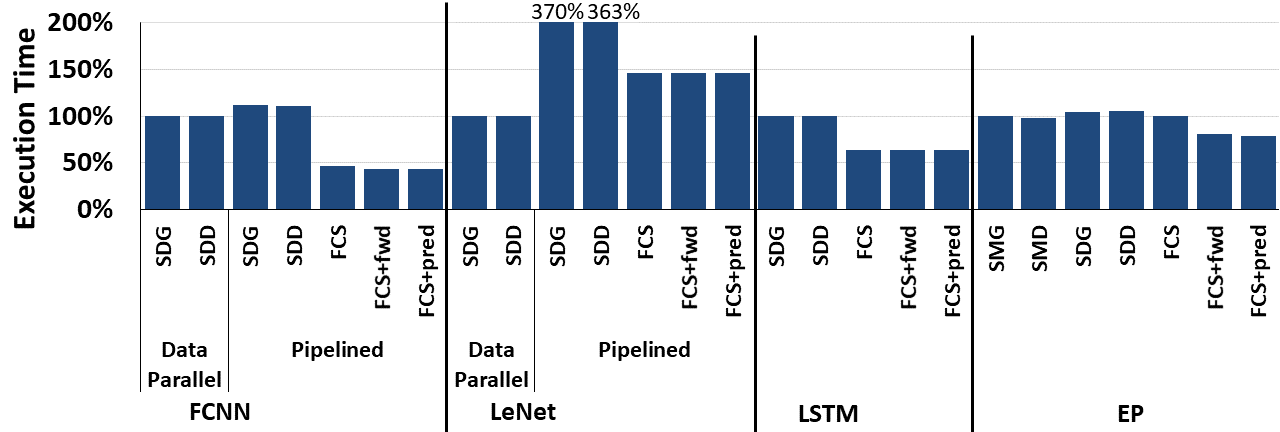} \\[\abovecaptionskip]
    \vspace{-1ex}
    \small (a) Execution time normalized to non-pipelined SDG
    \label{fig:fcnn-cycles}
  \end{tabular}
  
  \vspace{\floatsep}
  
  \begin{tabular}{@{}c@{}}
    \includegraphics[width=0.45\textwidth]{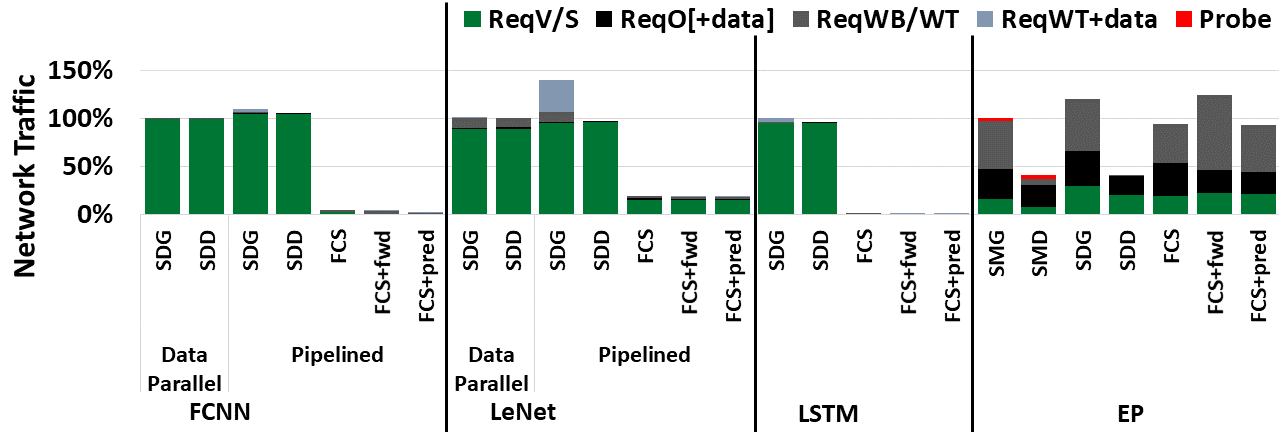} \\[\abovecaptionskip]
    \vspace{-1ex}
    \small (b) Network traffic normalized to non-pipelined SDG
    \label{fig:fcnn-nw}
  \end{tabular}
  
  \caption{\fontsize{9pt}{11pt} Execution time and network traffic for applications.
  }\label{fig:pipe-results}
  \vspace{-4ex}
\end{figure}

\vspace{-1ex}
Figure~\ref{fig:pipe-results} compares the execution time and network traffic for FCNN, LeNet, 
LSTM, and EP.
FCNN, LeNet, and LSTM are executed on the GPU, and the choice of CPU coherence strategy has no impact on results. 
We therefore consider only the SDG and SDD static coherence configurations.
For FCNN and LeNet, we compare data-parallel and pipeline-parallel implementations (FCS is only evaluated with the pipelined versions because data parallel implementations don't benefit from fine-grain coherence specialization).
EP runs on both CPU and GPU cores, so all static coherence configurations are compared (similar to the microbenchmarks). 


\subsubsection{FCNN}
\label{subsubsec:eval-pipe-FCNN}

The data parallel configurations exploit reuse in feature vectors and require no synchronization, but they are unable to exploit reuse in weight data since this will be evicted before it is reused.
Static coherence configurations that use pipelining increase execution time by 12\% relative to the data parallel implementations because they are unable to exploit reuse in feature values or weight values. 

When fine-grain coherence specialization is enabled (FCS, FCS+fwd, FCS+pred), pipelined FCNN obtains ownership for reads to layer-specific weight values and is able to exploit cache reuse in this data,
reducing execution time by 54\% and network traffic by 96\% relative to data parallel SDG.
Write-through Forwarding (FCS+fwd) additionally reduces execution time by 7\% and network traffic by 14\% relative to FCS by enabling producers of feature data and synchronization variables to forward them to the consumer.
Owner Prediction (FCS+pred) further reduces network traffic by 53\% relative to FCS+fwd by avoiding LLC lookups for the minimal remaining inter-device communication required in
FCS+fwd (as with FlexO/WT and Prod-Cons,\svac{Prod-Cons - do a search. Also be consistent in whether a slash is used between O/WT and F in Flex is capital etc. throughout the paper. The example figure showing all the arrays etc is inconsistent} execution time is minimally affected because GPU stores can tolerate latency).

\subsubsection{LeNet}
\label{subsubsec:eval-pipe-FFNN}



\hl{
As with FCNN, the 
weight data for the convolutional and fully connected layers (which make up a significant proportion of data accesses) will be evicted before they can be reused in a data parallel LeNet implementation.
}
\hl{
LeNet's pipelined implementation faces one significant challenge that did not exist for FCNN: pipeline imbalance.
This combined with an inability to reuse weight data across synchronization causes pipelined SDG and SDD to increase execution time by over 3x relative to data parallel SDG. 
By exploiting reuse for weight matrix reads, FCS 
eliminates much of the execution time overhead of the static pipelined implementations and reduces network traffic by 82\% relative to data parallel.
Relative to the fastest data parallel implementation, however, pipelined FCS+pred increases execution time by 46\% due to pipeline imbalance, which FCS is unable to address.
Even so, a pipelined FCS implementation is reduces the end-to-end latency of a single iteration by 41\% relative data parallel (which uses only a single device to execute each input), which may be important for latency-sensitive workloads.
}

\hl{
}
\subsubsection{LSTM}
\label{subsubsec:eval-pipe-rnn}


\hl{
Compared to FCNN, weight matrices make up an even larger proportion of total data accesses in an LSTM layer.
By obtaining ownership for this data, FCS almost entirely eliminates cache misses for LSTM, reducing network traffic by 99\% and execution time by 36\% relative to SDD.\svac{Confirming that lstm sdd is a pipelined implementation?}
Write-through forwarding (FCS+fwd) additionally reduces the network traffic by 13\% relative to FCS, and owner prediction (FCS+pred) further reduces the network traffic by 30\% relative to FCS+fwd by avoiding LLC lookups for the minimal remaining inter-device communication required in FCS+fwd. 
}

\subsubsection{EP}
\label{subsubsec:eval-EP}
In EP, both CPU and GPU exhibit high locality of shared data that is read and written with high reuse during the various execution stages.
As such, during GPU phases, static coherence configurations which use DeNovo for GPU (SMD, SDD) exploit inter-kernel reuse of written data and exhibit lower GPU execution time than GPU coherence for GPU (SMG, SDG).
However, this also results in reduced CPU performance for SMD and SDD, as the latency sensitive CPU must either read data or revoke ownership from a remote L1. \svac{As a result, the net effect is that both static DeNovo and GPU coherence show the same total performance.}
With FCS, the CPU gains ownership of the heavily accessed data, prioritizing latency sensitive CPU accesses, while the GPU uses write-through stores.
Without forwarding, FCS sees no benefit or improved reuse, since GPU writes must revoke the CPU's ownership.
However, FCS+fwd forwards GPU writes to the CPU owner and reuse increases for latency-sensitive CPU accesses, resulting in 17\% lower execution time versus the fastest static configuration, albeit at the cost of a 207\% increase in network traffic due to the additional GPU misses and the indirection incurred by write-through forwarding.
Using ownership prediction reduces this overhead, enabling a 20\% reduction in execution time with a 130\% increase in traffic relative to the fastest static configuration.

\hl{
}



\vspace{-1ex}
\section{Conclusion}
\label{sec:conc}
\vspace{-1ex}

As hardware adapts to address the growing memory bottleneck, efficient coherent data movement is increasingly important.
Existing heterogeneous hardware and software miss out on many opportunities for communication optimization because they are designed assuming static coherence protocols which are often complex and difficult to extend.

\johnhl{
Fine-grain coherence specialization enables more flexible and efficient hardware-software co-design, allowing future heterogeneous software to leverage data movement optimizations in hardware.
We present a methodology that empowers heterogeneous devices to mix and match specialized coherence request types based on individual access properties, including two new coherence specializations for data forwarding.
We show that the new specializations can be implemented without adding significant complexity to the underlying protocol as long as a simple and flexible protocol like Spandex is chosen as a baseline.
We then demonstrate the performance gains these optimizations can deliver for a set of microbenchmarks and applications. 
By giving software more control over data movement, 
we greatly improve cache efficiency, 
reducing execution time by up to 61\% and network traffic by up to 99\%.
}
Looking forward, this focus on flexible coherent data movement can improve cache efficiency, motivate new cross-stack coherence optimizations, and enable more efficient programming patterns for emerging heterogeneous workloads.


\section{Acknowledgements}
This work is supported in part by the National Science Foundation under grant CCF 16-19245, by DARPA through the Domain-Specific System on Chip (DSSoC) program, a Google Faculty Research award, and by the Applications Driving Architectures (ADA) Research Center, a JUMP Center co-sponsored by SRC and DARPA.

{\footnotesize
\bibliographystyle{IEEEtranS}
  \bibliography{hetero}
}

%

\end{document}